# SOLAR DIAMETER WITH 2012 VENUS TRANSIT: HISTORY AND OPPORTUNITIES

Costantino Sigismondi

ICRA, International Center for Relativistic Astrophysics   sigismondi@icra.it

Abstract. The role of Venus and Mercury transits is crucial to know the past history of the solar diameter. Through the W parameter, the logarithmic derivative of the radius with respect to the luminosity, the past values of the solar luminosity can be recovered. The black drop phenomenon affects the evaluation of the instants of internal and external contacts between the planetary disk and the solar limb. With these observed instants compared with the ephemerides the value of the solar diameter is recovered. The black drop and seeing effects are overcome with two fitting circles, to Venus and to the Sun, drawn in the undistorted part of the image. The corrections of ephemerides due to the atmospheric refraction will also be taken into account. The forthcoming transit of Venus will allow an accuracy on the diameter of the Sun better than 0.01 arcsec, with good images of the ingress and of the egress taken each second. Chinese solar observatories are in the optimal conditions to obtain valuable data for the measurement of the solar diameter with the Venus transit of 5/6 June 2012 with an unprecedented accuracy, and with absolute calibration given by the ephemerides.

Fruitful observations can be obtained also by amateur astronomers, by following the instructions in this paper. All ground-based observations designed to achieve this goal are warmly welcome to be analyzed by the author, presently visitinig the Huairou Solar Station of National Observatory of China for observing that transit. Finally a miminal observational schedule is suggested.

## 1 The method of eclipses

I. I. Shapiro in 1980[1] used records of transits of Mercury to recover the past history of the solar diameter.[2] Further studies seems to confirm the constancy of the diameter within the errorbars. Measurements made with different instruments, under perfect observing conditions, as in the case of Gambart and Bessel in 1832 yield different transit times, and different diameter of Mercury and, consequently, a different diameter of the Sun.[3, 4]

The determination of the planetary diameter is subjected to the Point Spread Function of the telescope matching with the Limb Darkening Function of the Sun,[5] and, in the case of Venus, there is also the atmosphere, already discovered by M. V. Lomonosov in 1761.[6]

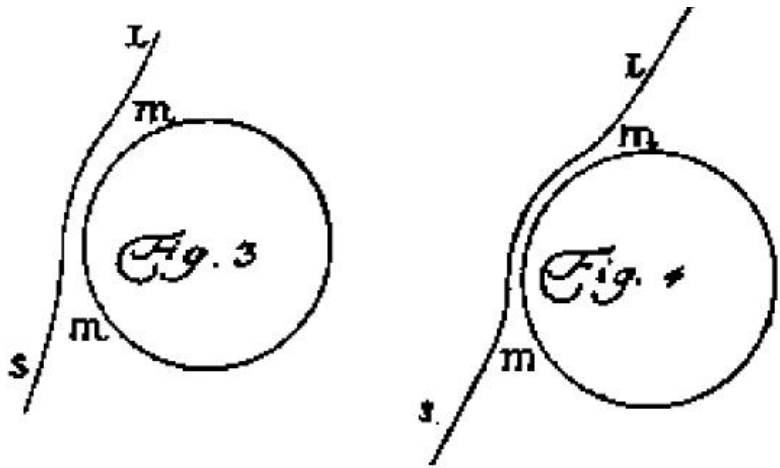

Fig. 1. Two stages of the egress of Venus in 1761 as observed by M. Lomonosov at the St. Petersburg Observatory [6].
He was the first to observe evidences of the presence of Venus atmosphere, and to predict its high density because of the refractive phenomenon.

Nowadays the chord draft on the solar limb by the planet's disk can be recovered by photos, in conditions not affected by black drop phenomenon.[7, 8] The time in which the chord is zero, when the black drop is maximum, can be extrapolated from UTC labeled photographs made each second around the intermediate stages of ingress and egress. After corrections for Earth's atmospheric refraction the ephemerides can be used to recover the solar diameter by comparison with the observed ingress and egress times.

The opportunity given by the forthcoming transit of Venus of 5/6 June 2012 and the one of Mercury of May 9, 2016 has to be exploited to measure the solar diameter with unprecedented accuracy. The studies on the Venus aureole[6, 9] if done with UTC synchronized high-resolution photos, can be useful to do solar diameter measurements, once the location of the observations are known with GPS coordinates.

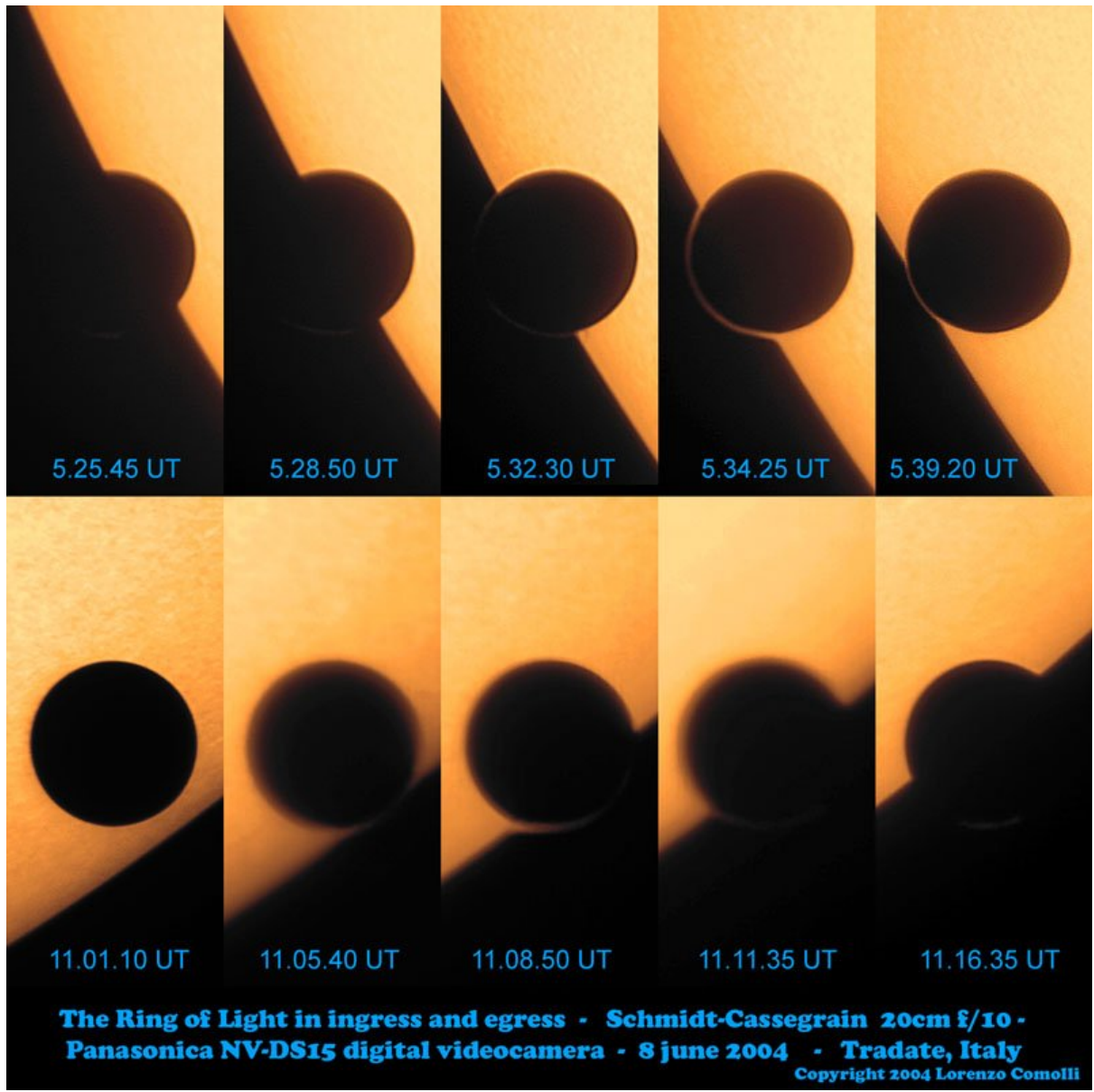

Fig. 3. Sequence of ingress/egress of last Venus transit of 2004 by L. Comolli. The required frequency to measure the solar diameter up to 0.01" of accuracy is one image per second. The refractive arc is visibile at both phases of the transit.

## 2. Method: one image per second during 2012 ingress/egress

We study the position of the center of the planet with respect to the inflexion point of the limb darkening function of the Sun (the solar limb). By using the circular fits to the undistorted part of the image we avoid the black drop effect. We can use two reference instants of comparison with ephemerides:
• the time when the center of Venus crosses the solar limb.
• the time when the chord drawn by Venus and the solar limb becomes zero.[8]
The first method can be applied also to the external aureole[9] produced by the refraction in the upper atmosphere of Venus, and this method is independent on the thickness of the atmosphere, because it uses the center of the planet. Similarly the determination of the center of the planet is less affected by the black drop phenomenon with respect to the second method of the length of the chord.

The method of the chord has been tested with 50 images made each minute, 25 at ingress and 25 at egress, in the H alpha line by Anthony Ayomamitis with a 16 cm apochromatic refractor near Athens during the Venus transit of 2004: the accuracy without refraction correction has been 2.6 s at the egress and 8.1 s at the ingress (with the Sun low near the horizon); the final accuracy on the whole solar radius was 0.38 arcsec.

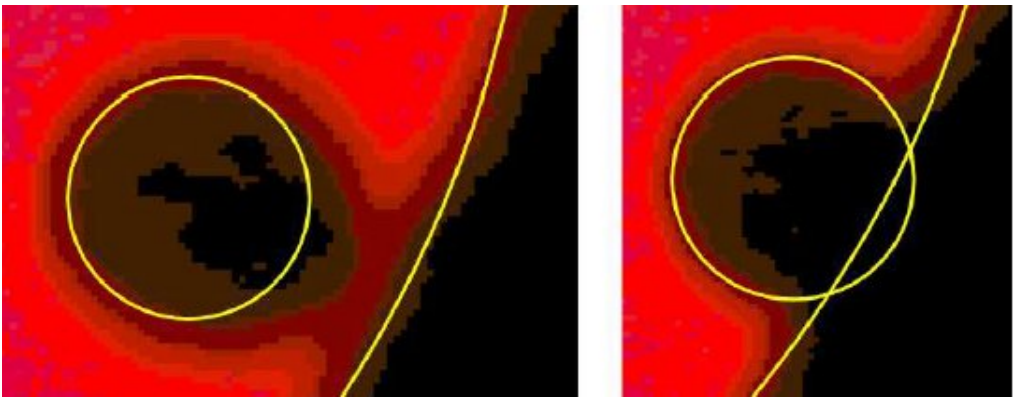

Fig. 2. Two stages of the egress of Venus in 2004 (H alpha images of Anthony Ayomamitis, Athens, details). The circular profiles of Venus and Sun are fitted to the undistorted part of the image.

An improvement on the final accuracy is expected with a 1 s photo sampling (60 times larger; accuracy $0.38/\sqrt{60} = 0.05$ arcsec), and a further improvement will come from the atmospheric refraction corrections and from the determination of the center of the planet, instead of the chord's length. The ultimate level of accuracy below 0.01 arcsec is the goal of this measurement: the most precise achievable with ground based methods.
Space observations will be also exploitable.[10]

## 3. Focus on practical questions

### 3.1 What is the logarithmic derivative of the solar radius with respect to the luminosity?

the logarithmic derivative of Luminosity L vs radius R is another way to define the percentual increment Delta L/L as a function of Delta R/R

Since: $L = 4\pi\sigma T_{eff}^4 R^2$ , $R = \left(L/4\pi\sigma T_{eff}^4\right)^{1/2}$

and consequently, with T constant: dLogR/dLog L = 0.5
when we observe different values of dLogR/dLogL, and this is the case, it means that, for example, the emissivity is not constant...

### 3.2 Is the solar diameter different at different wavelengths?

According to the figure from Irbah et al. "Ground-based solar astrometric measurements during the PICARD mission", submitted to SPIE (2011) there are different values of the diameter with respect to different wavelengths, ranging 0.7 arcsec within the visible domain.[7, fig. 1.16]
Neckel in 1994 published on Solar Physics a range for similar wavelengths of 0.07 arcsec. H alpha diameters are about 1" larger than photospheric diameters, since the H alpha processes occur higher than photosphere level.

### 3.3 How to obtain the solar diameter during the transit of Venus?

The solar photosphere is not polished: there are corrugations of the order of 100 Km peak to peak, some 0.1 arcsec peak to peak. These peaks fluctuate with timescales of some minutes, due to the local speed of sound typical of such phenomena. Moreover when we observe the Sun from ground-based observatories we are subjected also to the atmospheric seeing, with timescales of 1/100 s and amplitudes of 1-2 arcsec, i.e. the equivalent of 700-1400 Km at the solar distance.
The algorithm will smooth these time effects over several minutes of the ingress/egress phases, by fitting two circles in relative motion: the smaller one is the dark profile of Venus, and the larger one is the solar limb. The time of internal/external contacts of Venus with the solar limb will be identified from such fits. The chord drawn by Venus above the solar surface is proportional to the actual solar diameter; provided good ephemerides for representing this relative motion.
That's why we expect a final accuracy around 0.01 arcsec in the solar diameter, corresponding to less than 1s accuracy in the determination of such fitted times.

## 3.4 How and with which instruments we will observe the transit of Venus to measure the solar diameter?

Video sequences with regular cadence are required for these measurements. The rate of one photo each 2 or 5 seconds is recommended for all the durations of the ingress/egress phases.

Since the transit will last about 6 hours 40 minutes, 26400 s; a second of accuracy corresponds to 1/26400 of the solar diameter i.e. Less than 0.1 arcsec. The statistical accuracy can increase of another factor of 10 by using at least 100 images to determine each instant. That's why we expect 0.01 arcsec of final accuracy.

The photo have to be perfectly in focus, and with Venus at the center of the field of view in order to avoid optical aberrations. The filters used have to be communicated as well.

## 3.5 The importance of time synchronization of watches

 timing accuracy is crucial: the synchronization with UTC of the CCD/Cameras has to be checked, especially in case of single event observations, like in Europe, where the transit egress only will be visible. Since the solar diameter will be measured by the observed duration of the transit, the stability of relative timing is required over at least 5 hours at the temperature conditions of the observation. Usually a drift of 0.1 s per hour is normal for T=20 or 30°C: it has to be known in advance if observations of BOTH contacts are possible only with internal camera clock reference.

## 4 Suggested observational schedule

Try to make regular photos of the transit ingress and egress phases for the solar diameter measurement, i.e. from external to internal contacts of Venus with the solar limb.

With a photo rate of 1 per second a final accuracy of 0.01 arcsec is expected

                1 each 6 seconds the accuracy is 0.03 arcsec
                   1 each 1 minute the accuracy is 0.3 arcsec.

If you can't see both ingress and egress, please provide images with time synchronized with UTC as good as you can. If you see both ingress and egress try to save either UTC time and computer time.

Try to exploit the maximum resolution of your telescope plus CCD camera, by obtaining high quality images and well on focus. Concentrate yourself on the region of the ingress/egress, instead of obtaining a full disk image.

You can make a video of a good projected image on a white screen in a dark camera as well, if you have not CCD in your telescope. Remember to record also a synchronized watch during the same video.

Acknowledgments: Thanks to Anthony Ayomamitis for the images of the Venus transit of 2004. Despite of the worldwide campaign to observe this lifetime phenomenon nobody else (amateurs and observatories) yet published a sequence of chronodated images useful to measure the solar diameter by using the transit of Venus. Thanks to Patrick Rocher (IMCCE) for a fruitful discussion on the ephemerides. Thanks to Serge Koutchmy who pointed me out the paper on Lomonosov observation.